# On- and off-chain demand and supply drivers of Bitcoin price

Pavel Ciaian[a], d'Artis Kancs[a] and Miroslava Rajcaniova[b,c]

[a]European Commission, Joint Research Centre, Ispra, Italy; [b]Slovak University of Agriculture, Nitra, Slovakia; [c]University of West Bohemia, Pilsen, Czech Republic

**Abstract**
Around three quarters of Bitcoin transactions take place off-chain. Despite their significance, the vast majority of the empirical literature on cryptocurrencies focuses on on-chain transactions. This paper presents one of the first analysis of both on- and off-chain demand- and supply-side factors. Two hypotheses relating on-chain and off-chain demand and supply drivers to the Bitcoin price are tested in an ARDL model with daily data from 2019 to 2024. Our estimates document the differential contributions of on-chain and off-chain drivers on the Bitcoin price. Off-chain demand pressures have a significant impact on the Bitcoin price in the long-run. In the short-run, both demand and supply drivers significantly affect the Bitcoin price. Regarding transactions on the blockchain, only on-chain demand pressures are statistically significant – both in the long- and short-run. These findings confirm the dual nature of the Bitcoin price dynamics, where also market fundamentals affect the Bitcoin price in addition to speculative drivers. Bitcoin whale trading has less significant impact on price in the long-run, while is more pronounced contemporaneously and one-period lag.

**Disclaimer:** This work was supported by the Slovak Research and Development Agency under the contract No. APVV-22-0442 and by the Vega Agency under the project No. VEGA 1/0225/22. The conceptual framework of this paper is based on Ciaian et al. (2026). The authors are solely responsible for the content of the paper. The views expressed are purely those of the authors and may not in any circumstances be regarded as stating an official position of the European Commission.

# 1 Introduction

Crypto-currencies are exchanged between owners in many different ways. Depending on the platform used, crypto transactions can be regrouped into two broad types: "on-chain" and "off-chain", where "chain" refers to a sequence of blocks. Most of the existing crypto-currency literature has investigated "on-chain" transactions – mainly due to data availability considerations. In the same time, most of the crypto-coins are being exchanged "off-chain" (Tierno 2023; ESMA 2024). Their impact on crypto-currency prices is considerably less studied (Cerutti et al. 2024). The present study attempts to fill this gap by investigating how both on- and off-chain demand- and supply-side factors relate to crypto-currency prices.

"On-chain" crypto-asset transactions take place directly on the blockchain, they involve the transfer of crypto-coins from one digital wallet address to another and are recorded on the distributed ledger – blockchain. In distributed ledger technologies, data are structured into blocks and each block contains a transaction or bundle of transactions. The transaction validation is based on a consensus mechanism (Proof of Work in the case of Bitcoin), decentralisation (without intermediaries or a central authority) and cryptography (public-private key encryption), ensuring trustiness (no need for users to trust a central authority or intermediary), security (they cannot be altered once recorded), and transparency (all transactions are publicly visible and traceable) (Mount 2020; Soares 2022; Tierno 2023; ESMA 2024). On-chain transactions include, among other, transactions related to the purchase of services and goods (e.g., transaction from wallet to wallet), transactions related to decentralised finance (DeFi) (e.g., decentralised exchanges (DEXs), lending/borrowing) and other blockchain-based economic activities (e.g. gaming, supply chain management and traceability, non-fungible tokens).

The excessive on-chain transaction data growth – blockchain bloat – presents an increasingly pressing challenge for blockchain (Alzoubi and Mishra 2024). As one solution, an increasing share of crypto-asset transactions take place "off-chain". Off-chain transactions refer to all other crypto ownership changes that are not recorded on the blockchain. Typically, in an off-chain transaction, the legal ownership of a crypto asset changes, but it remains associated with the same digital wallet (for example, the wallet of a cryptocurrency centralised exchange (CEX)); the ownership change is not recorded on the blockchain. Instead, off-chain transactions are recorded on centralised ledgers or private order books of intermediaries such as crypto exchanges, custodial wallets, and financial institutions. Survey data of Blandin et al. (2020) show that off-chain transactions, both in terms of volumes and numbers, continue to be dominated by fiat-crypto-asset trades (and vice-versa), meaning that users primarily interact with 'gateway' service providers, such as exchanges, to enter and leave the crypto-asset ecosystem. The most prominent examples of off-chain transactions include transactions done on CEXs such as Binance or Coinbase.[1] Fiat-crypto transactions make up most of exchanges' trades, both in terms of trading volumes and transaction numbers (Blandin et al. 2020).

The main advantages of off-chain transactions are lower costs, faster execution of transactions and the ability to address privacy concerns (anonymity) (Mount 2020; Soares 2022; Tierno 2023; Alzoubi and Mishra 2024; ESMA 2024), as off-chain transactions offload activity from the blockchain. Off-chain transactions include, but are not limited to, transactions on CEXs related to the buying and selling of crypto coins, transactions for the purchase of services and goods (e.g. on the Lightning network), and financial transactions (e.g. lending/borrowing, margin trading). Off-chain transactions also include crypto-asset ownership changes on decentralised layer 2 protocols. Layer 2 protocols are secondary decentralised networks built on top of a primary blockchain to achieve greater scalability (handling a high volume of trades), faster transaction times and lower transaction costs. They process transactions on layer 2 protocols, which are then aggregated for a final settlement on the main blockchain (layer 1) (Soares 2022; Tierno 2023; Alzoubi and Mishra 2024; ESMA 2024).

---

[1] Another example is payment channels such as the Lightning Network, where transactions are not immediately recorded on the main blockchain. Instead, users can make multiple transactions within the channel and only the final balances are settled on the main blockchain (Mount 2020).

Most existing empirical analyses of crypto-assets are primarily based on data generated by on-chain activity ("on-chain data") (Ciaian et al. 2016; Cerutti et al. 2024). Analyses using on-chain data are useful, among others, for understanding the level of activity in the blockchain-based Bitcoin economy. For example, the volume of on-chain transactions indicate the adoption level of the blockchain-based crypto economy. An increase in the transaction volume is associated with higher velocity network traffic, user base and trading activity, and more generally, trust in the blockchain-based crypto economy. In contrast, a decrease in the transaction volume may signal uncertainty – as perceived by users – and a lower level of the overall adoption of the blockchain-based economy. On-chain data also provide an understanding of the share of illicit activity using cryptocurrencies, which is estimated at 0.34% of the total transaction volume (Chainalysis 2024). On-chain analysis is further useful to deduce the value of crypto-assets being moved on-chain between real-world entities, demonstrating for instance that exchanges account for 90% of all funds sent by crypto-asset services (Blandin et al. 2020).

Despite their merits, layer one blocks on distributed ledgers do not contain information about off-chain sales, which are recorded on private order books of intermediaries such as crypto-exchanges or financial institutions. Given that on-chain data exclude purchases of crypto-assets with fiat currency, sales of crypto-assets for fiat currency and swaps between crypto-assets, they provide a partial – and likely biased – market-level picture of crypto-trading. To gain an understanding of these and other transactions on different blockchain layers, an analysis leveraging off-chain data is necessary. While the existing large and vibrant literature has looked at on-chain trading in crypto-assets, little empirical evidence exists on how investors trade off-chain, what are the aggregated market-level consequences and impacts on crypto-asset returns. The present study aims to fill this research gap by studying both on- and off-chain demand and supply drivers of the short run price.

## 2 Research Hypotheses

We examine Bitcoin transaction patterns and develop two hypotheses that are tested empirically using time-series mechanisms in the following. One hypothesis relates on-chain versus off-chain transactions to the Bitcoin price. The other hypothesis relates the trading of large coin owners to its price.

### 2.1 Off-chain driver Hypothesis

There are at least three reasons why crypto-currency trading fundamentals may differ between on- and off-chain transactions, and on-chain crypto-coin users and off-chain crypto-asset traders would respond differently to the same market signals, triggering a differentiated impact on crypto-prices. First, the evidence suggests that the main purpose of off-chain transactions is fundamentally different from on-chain crypto-asset transactions (Makarov and Schoar 2021; Feyen et al. 2022; Tierno 2023; Cerutti et al. 2024). Second, the underlying data moments, such as mean and variance, show significant structural differences between on-chain and off-chain transactions. Third, the share of off-chain transactions makes up the majority of the total Bitcoin transaction volume (Figure 1), suggesting that their impact on crypto-asset prices would be significant. In light of these structural differences between off-chain and on-chain transactions, the 'Off-chain driver Hypothesis' investigates the possible role of off-chain transactions in the BitCoin price formation.

Crypto-coin users and investors make their decisions regarding the specific blockchain layer to be used depending on the purpose, scale and frequency of their intended transactions: usually layer 1 for the purchase of services and goods and transactions related to decentralised finance, whereas layer 2 for fiat-crypto-asset trades (Mount 2020). The vast majority of on-chain transactions are not tied to economically relevant activities. According to Makarov and Schoar (2021), approximately 90% of on-chain Bitcoin transaction volume between 2017 and 2020 represents spurious transactions in which entities exchange Bitcoin among themselves (i.e. the equivalent of someone moving cash from one pocket to another). The remaining 10% of on-chain transaction volume represents genuine (real) transactions between different entities. Makarov and Schoar (2021) also find that about 80% of the on-chain genuine transaction volume is related to exchanges and trading desks (e.g. OTC brokers, institutional traders). A significant portion of the volume on exchanges is initiated by investors who hold Bitcoins off-exchange and move them just in time to trade (Hoang and Baur 2022). The majority of off-chain Bitcoin trading takes place on CEXs rather than DEXs, further hinting to a possibly differentiated impact of off-chain transactions on the Bitcoin price. According to CoinMarketCap.com, which tracks the performance of various cryptocurrencies, the DEX trading volume accounts for less than 1% of the total daily Bitcoin trading volume (CoinMarketCap 2024).[2] Off-chain trading on CEXs is usually perceived as speculative transactions, carried out with the aim of extracting gains from price movements or hedging against alternative investments (e.g. stocks, commodities) rather than sustaining economic activities such as purchase goods and services (Kukacka and Kristoufek 2023; ESMA 2024; Ozer et al. 2024). According to Hougan et al. (2019), 95% of Bitcoin transactions on CEXs are related to the purchase and sale of Bitcoins without an economic value, but mainly involve fake or wash trading. This implies that off-chain transactions tend to be primarily speculative in nature. In contrast, on-chain Bitcoin transactions (after excluding on-chain transactions with DEXs) are found to be more likely driven by market fundamentals (Ciaian et al. 2016), as on-chain Bitcoin trading is significantly less common.

Second, according to Cerutti et al. (2024), on-chain transactions differ, on average, significantly from off-chain transactions. For example, on-chain transactions are, on average, significantly larger than off-chain transactions (LocalBitcoins). The average transaction size amounts to 13.3486 Bitcoin on the blockchain compared with 0.0178 Bitcoin off-chain. Likewise, at the same Bitcoin price, the maximum transaction amounts to US$300,000,000 on the blockchain compared with US$1,875,000 off-chain. The difference in off-chain and on-chain transaction sizes and distribution reflect distinct groups of market participants (Cerutti et al. 2024). The IMF evidence suggests that circumvention of capital flow restrictions and transfers of remittances are major incentives behind cross-border off-chain transactions. The descriptive statistics presented in Table 3 confirm that the underlying data moments, such as mean and variance, show significant structural differences between on-chain and off-chain drivers. For example, on-chain Bitcoin demand side drivers – as captured by *On-chain BTC transactions* – are fundamentally different from off-chain Bitcoin demand side drivers – as captured by *Bank netflow*, *Bank reserve*, and *Fund volume*. Importantly, differences between on-chain and off-chain drivers are structural, as the underlying trading motivation and purpose is different (Cerutti et al. 2024).

---

[2] The main reason is that the Bitcoin ecosystem is relatively underdeveloped compared to other cryptocurrencies such as Ethereum, as blockchain lacks the programmability that would support smart contracts. This limitation constrains the development of decentralised applications (dApps) such as DEXs in contrast to Ethereum, which is the leading blockchain platform in the dApp economy (Leiponen et al. 2022).

**Figure 1. Global monthly BitCoin on-chain and off-chain transaction volume 2020-2024**

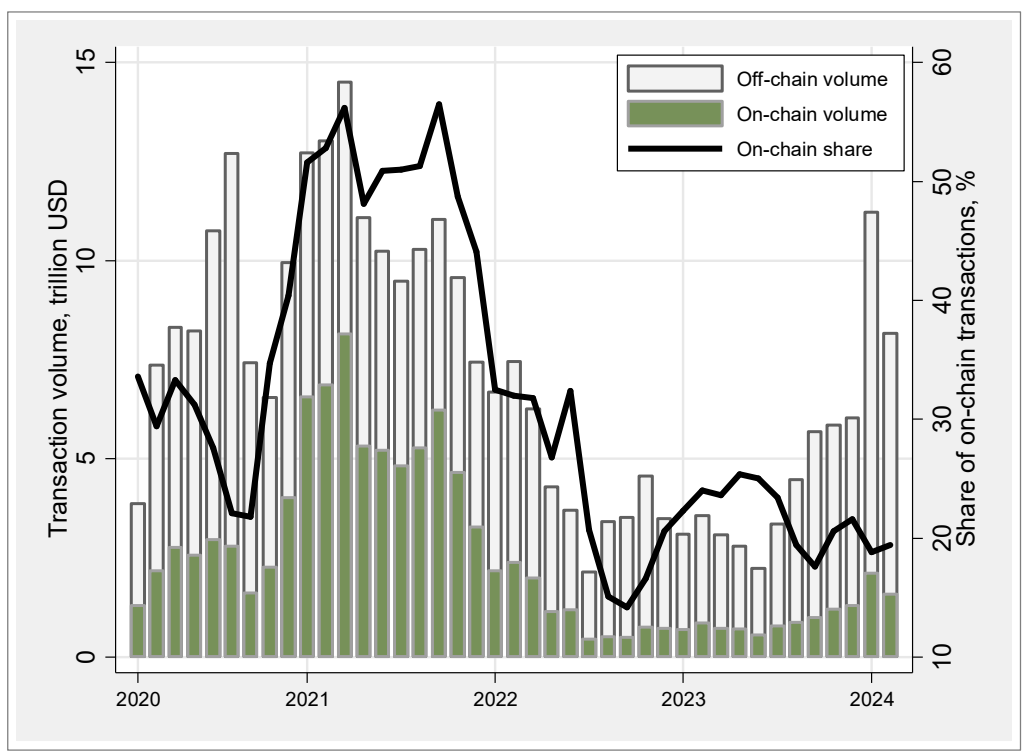

Source: Authors' computations based on on-chain transaction volume data from theblock.co and off-chain trading volume data from ccdata.io. Notes: BitCoin transaction volume is measured in trillion USD current prices on the left Y axis.

Third, depending on the share of off-chain transactions vis-à-vis on-chain in the total crypto-asset transaction volume, the impact of off-chain drivers on crypto-asset prices may be significant. To contextualise the scale of off-chain transactions compared to on-chain crypto-asset transactions, Figure 1 plots the global monthly BitCoin on-chain transaction volume (dark bars) and off-chain transaction volume (light bars), measured in trillion USD current prices on the left Y axis – based on data from the theblock.co and ccdata.io. The on-chain volume is defined as the volume of digital coins transferred; all transactions written on the underlying blockchain. Figure 1 reveals: (i) the ratio of on-chain to off-chain transactions is varying considerably over time, the minimum of on-chain transaction share falling below 15% while the maximum raising to above 55%. (ii) During the last two years, the on-chain – off-chain transaction ratio has stayed comparably stable – between 15% and 25%. On the right Y axis, Figure 1 measures the share of on-chain transactions in the total BitCoin transaction volume – solid black line. (iii) Overall, the total off-chain volumes appear significantly larger than on-chain transactions with our estimates suggesting the approximate ratio of off-chain to on-chain volume being roughly 5:1 during the last two years.

This estimated ratio is close to Feyen et al. (2022), deriving a similar ratio of off-chain to on-chain volume being roughly 6:1. Makarov and Schoar (2021) estimate that since 2015, approximately 75% of the total real Bitcoin volume has occurred off-chain, i.e. through exchanges or exchange-like entities such as on-line wallets, OTC desks, and large institutional traders, implying a ratio of off-chain to on-chain volume 4:1. Between 2017 and 2020, the weekly on-chain genuine transaction volume typically varied between 50 and 160 thousand, while the weekly off-chain transaction volume varied between 100 and 300 thousand Bitcoins.

In summary, the existing evidence in the literature along with the descriptive statistics presented above suggest that off-chain transactions and genuine on-chain transactions transferred to exchanges for off-chain trading account for the vast majority of Bitcoin ownership changes. Both demand- and supply-related drivers of off-chain transactions are potentially fundamental determinants of the Bitcoin price. Differences in the relative importance of different types of Bitcoin transactions (on-chain versus off-chain) are expected to result in a differentiated impact on the Bitcoin price.

**Hypothesis 1:** *Off-chain demand and supply drivers are expected to have a significant impact on the Bitcoin price.*

**2.2 Whale Hypothesis[3]**

The distribution of Bitcoin ownership is highly unequal, a significant portion of the total Bitcoin supply is held by a small group of individuals or entities, often referred to as "crypto whales". One of the key metrics used to measure the ownership concentration is the number of wallets (column 3 in Table 1) holding a given range of Bitcoin and as a percentage of total Bitcoin holdings (column 4 in Table 1).

**Table 1. BitCoin ownership distribution: entity count versus entity balance, 2023**

|  | Entity share | Entity count | Balance, share | Balance, Million |
|---|---|---|---|---|
| **>100BTC** | **0.043%** | **13,840** | **48.1%** | **9.26** |
| >5000BTC | 0.001% | 190 | 14.3% | 2.75 |
| 1000-5000BTC | 0.004% | 1,450 | 14.0% | 2.69 |
| 500-1000BTC | 0.007% | 2,200 | 8.7% | 1.68 |
| 100-500BTC | 0.031% | 10,000 | 11.1% | 2.14 |
| **1-100BTC** | **2.6%** | **832,000** | **24.1%** | **4.65** |
| 50-100BTC | 0.2% | 12,000 | 4.6% | 0.89 |
| 10-50BTC | 0.2% | 80,000 | 9.0% | 1.74 |
| 1-10BTC | 2.2% | 740,000 | 10.5% | 2.02 |
| **<1BTC** | **97.3%** | **32,000,000** | **6.5%** | **1.25** |
| Miners | 0.1480% | 50,000 | 9.5% | 1.83 |
| Exchanges | 0.0001% | 20 | 11.7% | 2.26 |

Source: Authors' computations based on supply distribution of Bitcoin data from ChainExposed.com and glassnode.com. Notes: To improve the accuracy and precision for measuring Bitcoin ownership, and isolate large entities such as exchanges or ETF products which represent large collective user-bases, multiple addresses of a single entity owner are collated and grouped with entity-adjustment clustering algorithms.

One approach to gain insights in the Bitcoin ownership distribution is to use network addresses' data, e.g. from bitinfocharts.com. The top 1% of Bitcoin accounts own over 90% of the total supply, 2% of accounts control 95% of all Bitcoin (Bitinfocharts 2024). A single user possesses 0.78% of all Bitcoins in circulation, while the cohort of top 100 hold 13.52% of all Bitcoins. The network addresses approach has caveats, however (Glassnode 2023). On the one hand, not all Bitcoin addresses can be treated equal. For example, an exchange address holding the funds from many users needs to be distinguished from an individual's self-custody address. On the other hand, a Bitcoin address is not an "account". One user can control multiple addresses, whereas one address can hold the funds from multiple users.

The application of heuristics and entity-adjustment clustering algorithms to the network addresses data – which collates and groups multiple addresses having a single entity owner – allows us to create an upper bound for the actual number of network participants. For example, Glassnode (2023) has analysed the distribution of Bitcoin across entities of different sizes, taking into consideration addresses that belong to exchanges and miners. Broadly confirming Bitinfocharts (2024) statistics, the Bitcoin supply held by the institutional investor cohort has further increased compared to 2021, suggesting an increase in large investors. As shown in Table 1 (row 2), 190 largest crypto-whales each holding 5000 coins or more together own 2.75 Million coins (14.3% of the total Bitcoin supply). Compared to 2021, their share has increased by one percentage point from 2.47 Million (13.3%) (Glassnode 2023).[4] This highlights a significant consolidation of Bitcoin ownership by large mostly institutional investors. While whales continue to accumulate Bitcoin, small "mom-and-pop" investors represent a decreasing ownership share. Sai et al. (2021) analysis of on-chain data confirm an increase in the concentration of the institutional investor cohort: in 2021 0.16% of the largest Bitcoin addresses held over 82% of all Bitcoins in circulation, while the smallest 85.5% of addresses held 0.70% of all Bitcoins. 74% of Bitcoin owners hold less than around 0.01 worth of Bitcoin According to Glassnode (2023), at the bottom end of the ownership's distribution, 97.3% of all Bitcoin owners (ca. 32 Million mom-and-pop investors) hold less than 1 Bitcoin per owner and 6.5% of the total Bitcoins. The growing concentration of Bitcoin ownership exposes small retail investors to higher risk and vulnerability as, on average, small investors face higher volatility than large investors (Gabaix et al. 2006; Choi and Chhabria 2012).

---

[3] Throughout the paper, we refer to a 'crypto whale' as an entity that owns a significant amount of cryptocurrency. Unlike an average trader, whales can operate on a massive scale, influencing the market's volatility and returns.

There are at least two reasons, why the Bitcoin accumulation and trading patterns of crypto-whales could disproportionally affect the Bitcoin price (Rose 2023). First, investors holding large crypto-asset portfolios can influence the supply and demand of Bitcoin in circulation and cause price fluctuations through their trading decisions. Specifically, the engagement of whales in crypto-markets can have a two-fold effect. On one hand, their confidence and accumulation can attract more crypto-buyers and drive the price higher. On the other hand, their potential selling activities can cause sharp drops in price (ESMA 2024). Historically, whale activities have often preceded market shifts (see Figure 2). Whales' accumulation patterns signify confidence in Bitcoin's long-term value, potentially hinting at bullish market trajectories. Conversely, substantial sell-offs by crypto-whales have often foreshadowed bearish trends. According to Glassnode (2023), Bitcoin whales have long been pivotal in shaping market directions. Further, trading decisions of whales are followed and coat tailed by many small retail investors, which can magnify the initial market-liquidity effects. Particularly in periods with high market uncertainty and in the presence of costly information coattail investors copycat the actions of famous and successful investors. When many regular investors replicate the investment decisions of whales' trading, the coattailing can lead to a herding effect and magnify the initial market impacts of crypto-whales (Spyrou 2013; Merkley et al. 2024).

Second, crypto whales have the potential to influence market dynamics by trading strategically and exploiting market imperfections. Through a strategic trading, including pump-and-dump schemes, Bitcoin whales can affect market liquidity by accumulating/releasing large amounts of Bitcoins, thereby tapering/easing coin scarcity. Choi and Chhabria (2012) find that whale investors have engaged in front-running ahead of funds by buying ahead of the investing public, and with subsequent and strategic disclosure anticipate that opportunistic investors might flock behind, thereby bidding up prices. An increasing crypto-coin ownership's concentration makes it easier to manipulate market by carrying out a coordinated pump-and-dump strategy. Large Bitcoin holdings in a few hands are easier used to initiate whale-driven price fluctuations, causing increased volatility and uncertainty for smaller investors. Indeed, Glassnode (2023) report that Bitcoin whales face low volatility, even in periods of high activity. Additionally, the whales' ability to coordinate their actions could potentially sway market sentiment, impacting retail traders and investors alike. Gabaix et al. (2007) show that the trading activity of large investors can affect prices and trading volume, creating profit opportunities through strategic trading and exploiting market inefficiencies. Even in the absence of fundamental market news, whale-investors can leverage their market influence in an illiquid crypto-coin market. Gabaix et al. (2006) show how trading by individual large investors may create price movements that are hard to explain by fundamental news.

---

[4] In June 2024, Bitcoin has a circulating supply of 19.71 Million coins, the maximum supply will be 21.00 Million.

**Figure 2. Whale balances (entities with >1000BTC) and BitCoin price**

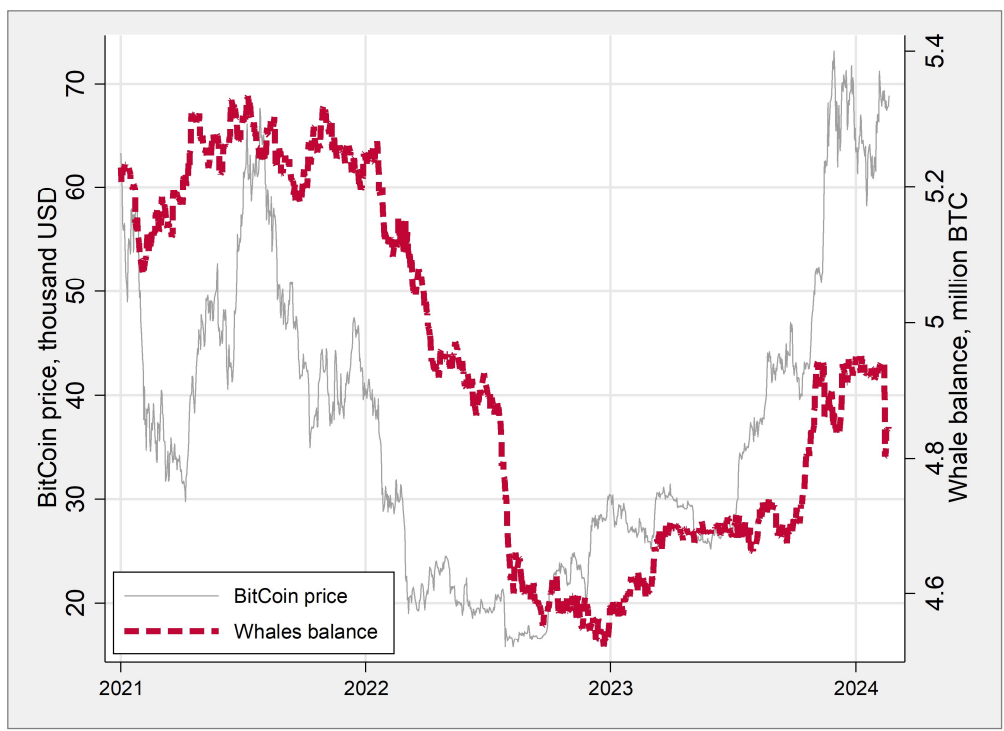

Source: Authors' computations based on supply distribution of Bitcoin data from ChainExposed.com and glassnode.com. Notes: To improve the accuracy and precision for measuring Bitcoin ownership, and isolate large entities such as exchanges or ETF products which represent large collective user-bases, multiple addresses of a single entity owner are collated and grouped with entity-adjustment clustering algorithms.

In summary, whales – large investors with significant Bitcoin holdings – control sizeable amounts of the total coins in circulation; the overall ownership of Bitcoin is characterised by a considerable skewness. The fat-tailed distribution of investor sizes generates a fat-tailed distribution of volumes and returns (Gabaix et al. 2006). The two channels through which crypto-whales affect the Bitcoin price are: (i) Whales-investors holding significant amounts of crypto-assets can influence the supply and demand of Bitcoin and may create price movements that are hard to explain by fundamental news. Prominent crypto-whales with many followers, large institutional investors and crypto exchange CEOs play pivotal roles in the crypto market, their strategic decisions, including significant investments and corporate treasury allocations, set precedents and impact crypto-asset market liquidity and trading choices of many coattail investors. (ii) Whales have the potential to influence market dynamics through strategic trading and exploiting market inefficiencies. By trading strategically and manipulating the actions of retail traders provides the possibility to influence liquidity and market dynamics, how trading by individual large investors may create price movements that are hard to explain by fundamental news.

**Hypothesis 2:** *The trading patterns of Bitcoin ("whales") are expected to significantly affect the Bitcoin price.*

## 3 Methodology

To identify a differentiated effect of on-chain and off-chain transactions and their associated supply and demand drivers on the Bitcoin price, we employ an autoregressive distributed lag (ARDL) model. The ARDL model is a versatile econometric tool widely used for analysing the relationship between financial time series data (e.g., Stoian and Iorgulescu 2020). Compared to other conventional cointegration techniques, the ARDL model is particularly advantageous in capturing both short-term and long-term effects, making it an ideal model for our inquiry in Bitcoin price drivers. Importantly, the ARDL method handles different lag lengths for various regressors. Further, a correctly specified ARDL model can address the issues of endogeneity and serial correlation concurrently (Pesaran and Shin, 1999), which is relevant in our context given that the testable hypotheses derived in the previous section include interdependent variables.

As usual, we start with investigating the presence of a long-term relationship between time series using the ARDL bounds test introduced by Pesaran et al. (2001). Unlike standard cointegration approaches, the ARDL method can be applied to time series that are stationary at levels I(0), first differences I(1), or cointegrated (Pesaran et al. 2001). Nevertheless, as Ouattara (2004) points out, the F-statistics provided by Pesaran et al. (2001) become invalid if I(2) variables are included in the model. To ensure no time series are integrated of order I(2) or higher, we use the Augmented Dickey-Fuller (ADF) test and the Phillips-Perron (PP) test to assess the stationarity of the data series and their first differences. The Akaike Information Criterion is used to decide about the optimal number of lags. In a general form, the ARDL (p, q) model reads as follows:

$$y_t = c_0 + c_1 t + \sum_{i=1}^{p} \phi_i y_{t-i} + \sum_{i=0}^{q} \beta_i x_{t-i} + \gamma z_t + u_t$$

where, $c_0$ and $c_1 t$ are intercept and a linear trend, respectively, y is the dependent variable (Bitcoin price), x is a vector of independent variables (demand and supply drivers related to different types of BTC transactions, macro-financial variables), p refers to the number of optimal lags of the dependent variable, q is the number of optimal lags for each explanatory variable and $u_t$ is a white noise error term. We may include a set of exogenous variables $z_t$ with predictive power to better explain the short-term deviations of y without impacting its equilibrium (Kripfganz and Schneider 2023).

The ARDL bounds testing technique is used to check for a long-term relationship. This involves calculating F-statistic and t-statistic and comparing them to critical value bounds. Pesaran et al. (2001) have suggested two types of critical values for a given significance level: one assumes all variables are I(1), while the other assumes all series are I(0). If the calculated F-statistic and t-statistic are below the lower bound, the null hypothesis of no long-term relationship cannot be rejected, indicating that an ARDL model in first differences without an error correction term should be estimated. If the F-statistic and t-statistic are found between the lower and upper bounds, the outcome is inconclusive. If the F-statistic and t-statistic cross the upper bound, the null hypothesis of no cointegration can be rejected. In this case, the error correction model to be estimated is (Hassler and Wolters 2006):

$$y_t = c_0 + c_1 t - \alpha(y_{t-1} - \theta x_{t-1}) + \sum_{i=1}^{p-1} \psi_{yi} \Delta y_{t-i} + \omega \Delta x_t + \sum_{i=0}^{q-1} \psi_{xi} \Delta x_{t-i} + \gamma z_t + u_t$$

where θ denote the long-run coefficients, ψ are short-run multipliers and α is the speed of adjustment of the dependent variable to a short-term shock, which indicates the speed at which the variables revert to their long-term equilibrium.

As suggested by Pesaran et al. (2001), we have conducted a series of diagnostic tests to validate the ARDL results, assuming normally distributed error terms, no serial correlation, no heteroscedasticity, and coefficient stability. The model specification and number of lags were determined based on these diagnostic tests, including the Breusch-Godfrey LM test and Durbin's alternative test for autocorrelation, the Breusch-Pagan/Cook-Weisberg test for heteroscedasticity, normality testing, and the cumulative sum test for parameter stability.

## 4 Data and variable construction

In the empirical estimates, we use daily data for the period 04/12/2019-25/01/2024. A detailed summary of the data used in the estimations and their sources is provided in Table 2. Table 3 provides descriptive statistics of the time series. In all specifications, our dependent variable is the price of Bitcoin, expressed in US dollars per Bitcoin.

In line with Hypothesis 1, we include several explanatory variables to proxy the demand and supply drivers of on-chain and off-chain Bitcoin transactions. Regarding the off-chain demand side, we consider three alternative variables: *Bank netflow*, *Bank reserve* and *Fund volume*. As these variables represent the demand side, it is expected that they will have a positive relationship with the price of Bitcoin. *Bank netflow* measures the net amount of Bitcoin flowing into and out of the digital asset banks that provide various financial services including lending, custody, staking, payment, synthesised assets (e.g. stablecoins or tokenised assets). This metric provides insights into the overall Bitcoin demand. A positive net flow suggests an increasing demand as more investors are depositing Bitcoin into trading platforms, potentially to hold in the long term. Conversely, if the net flow is negative, it suggests reduced demand as more investors are withdrawing Bitcoin from these platforms, potentially to hold in private wallets or for other purposes. *Bank reserve* represents the USD value of coins held by the digital asset banks and indicates their level of liquidity. A larger *Bank reserve* indicates greater liquidity and therefore demand potential with more funds readily available to facilitate transactions. *Fund volume* refers to the driver of off-chain transactions associated with Bitcoin-regulated funds such as trusts, ETFs and mutual funds. *Fund volume* measures the Bitcoin trading volume of regulated funds; its increase is associated with an increase in liquidity of regulated funds and implies an upward pressure on the Bitcoin price.

Regarding the off-chain supply side, we consider two alternative variables: *Exchange netflow* and *Exchange reserve*. These variables are expected to be negatively related to the Bitcoin price. *Exchange netflow* refers to Bitcoin supply drivers as it reflects the movement of Bitcoin into and out of exchanges, directly influencing the supply dynamics in the CEX market. An increase in *Exchange netflow* suggests that more Bitcoin is being moved to exchanges, which can be an indicator of potential selling activity and exert a downward pressure on the Bitcoin price. *Exchange reserve* refers to the total amount of Bitcoin held in CEXs. It gives an indication of the total supply of Bitcoin that is readily available for trading. A higher reserve means more Bitcoin is available on exchanges, which would potentially increase the volume of coins offered.

The variable *On-chain BTC transactions* proxies for on-chain demand side transactions of Bitcoin. It is calculated by subtracting the transactions flowing into or out of exchanges from the total on-chain Bitcoin transactions.[5] A higher value of this variable would indicate a higher intensity of not-exchange-related Bitcoin activities (non-DEX), which is expected to be positively related to the Bitcoin price.

The variables related to on-chain supply side considered in estimations include *Total supply* and *Coin days destroyed*. These variables are expected to be negatively related to the Bitcoin price. *Total supply* captures the total issued (minted) Bitcoins. It measures the cumulative amount of all Bitcoins that have been created since the inception of Bitcoin and thus provides a measure of the circulating supply of Bitcoins. *Coin days destroyed* is calculated by taking the number of Bitcoins in transaction and multiplying it by the number of days since those coins were last spent. An increase in this variable suggests that long-term holders of Bitcoin (inactive coins) are liquidating their positions, potentially exposing their holdings to selling pressure.

In line with Hypothesis 2, we proxy the whale demand and supply drivers of off-chain Bitcoin transactions: *Bank whale netflow* and *Exchange whale netflow*, respectively. *Bank whale netflow* measures the net amount of the top 10 Bitcoin transactions flowing into and out of the digital asset banks. A positive net flow suggests an increased whale demand as more large investors deposit Bitcoin into these off-chain platforms. *Exchange whale netflow* reflects the movement of the top 10 Bitcoin transactions into and out of CEXs. An increase in this variable indicates that more whales are moving Bitcoin into exchanges, indicating a potential whale selling pressure on the Bitcoin price. The former variable is expected to have a positive relationship with the price of Bitcoin, while the latter is expected to have a negative relationship.

Following literature (Apergis 2024, Ozer et al. 2024), we include a series of control variables. Bitcoin is often considered as an investment asset, with potential investors weighing the expected benefits of investing in Bitcoin against other assets or using it as an inflation hedge (Ciaian et al. 2016; Sören 2023; Cong et al. 2024). As a result, macro-financial developments (e.g. stock market, inflation, gold price) are also expected to influence the price of Bitcoin, they also represent off-chain drivers (Apergis 2024; Ozer et al. 2024). For this reason we consider several macro-financial variables: Federal Funds Effective Rate (*DFF*), Market Yield on U.S. Treasury Securities (*DFII10*), Consumer Price Index (*CPIAUCSL*), Wilshire 5000 Price Index (*WILL5000PR*) and *Gold price*. *DFF* is the US Federal Reserve's main tool for influencing monetary policy. Changes in *DFF* reflect the US Federal Reserve's stance on monetary policy, with decreasing rates indicating an accommodative approach (expansionary) to stimulate economic activity and inflation, while increasing rates signal a more restrictive stance aimed at controlling inflation. *DFII10* represents the real yield or real interest rate on U.S. Treasury securities with a maturity of 10 years adjusted for changes in inflation as measured by the Consumer Price Index (CPI). *CPIAUCSL* is a measure of inflation of USD. WILL5000PR represents the market value of all American stocks actively traded in the United States. Gold price represents a commodity asset price – typically considered as a store of value by investors.

As Bitcoin is often regarded as a store of value asset (Yae and Tian 2024), a negative relationship between the Federal Funds Effective Rate (*DFF*) and the Bitcoin price is expected: the price of Bitcoin is expected to increase when monetary policy is expansionary (low *DFF*), while it is expected to decrease when monetary policy is contractionary (high *DFF*). Similar holds for *CPIAUCSL*. In the presence of higher inflation (higher *CPIAUCSL*), the Bitcoin price is expected to increase if Bitcoin is perceived as a store of value asset. Potential investors also often consider Bitcoin as an alternative investment opportunity among many other possible investment opportunities (such as stocks, treasury bonds). Given that Bitcoin competes with

---

[5] This is done using the *Fund Flow Ratio* which is calculated by dividing the total amount of Bitcoin flowing into or out of exchanges by the overall Bitcoin amount transferred across the entire Bitcoin network.

other financial assets for the attention of investors, it has to deliver a competitive expected return. The return arbitrage between alternative investment opportunities implies a positive price relationship between the price of Bitcoin and financial assets (i.e. *DFII10*, *WILL5000PR*) (Ciaian et al. 2018; Apergis 2024, Ozer et al. 2024). *DFII10* serves as a proxy for the risk-free investment alternative, while *WILL5000PR* represents higher-risk investment alternatives. *Gold price* is expected to be positively related with the price of Bitcoin, as it may represent an alternative investment opportunity, an alternative store of value or both

**Table 2. Variable description and data sources**

| Variable name | Variable description/formula | Source |
|---|---|---|
| *Dependent variable* | | |
| Bitcoin price | USD per Bitcoin | |
| *Off-chain demand side* | | |
| Bank netflow | Bank Netflow (Total) | cryptoquant.com |
| Bank whale netflow | Bank Whale Netflow (Top10) = (Bank Inflow (Top10)) – (Bank Outflow (Top10)) | Calculated based on data from cryptoquant.com |
| Bank reserve | Bank Reserve USD | cryptoquant.com |
| Fund volume | Fund Volume - All Symbol | cryptoquant.com |
| *Off-chain supply side* | | |
| Exchange netflow | Exchange Netflow (Total) - All Exchanges | cryptoquant.com |
| Exchange whale netflow | Exchange Whale Netflow (Top10) = (Exchange Inflow (Top10) - All Exchanges) - (Exchange Outflow (Top10) - All Exchanges) | Calculated based on data from cryptoquant.com |
| | *Not BTC specific* | |
| Exchange reserve | Exchange Reserve - All Exchanges | cryptoquant.com |
| *On-chain demand side* | | |
| On-chain BTC transactions | On-chain BTC transactions = [Tokens Transferred (Total)] * [1- (Fund Flow Ratio - All Exchanges) / 100] | Calculated based on data from cryptoquant.com |
| *On-chain supply side* | | |
| Total supply | Total Supply | cryptoquant.com |
| Coin days destroyed | Bitcoin Coin Days Destroyed (CDD) | cryptoquant.com |
| *Macro-financial variables* | | |
| DFF | Federal Funds Effective Rate, Percent, Daily, Not Seasonally Adjusted. | Federal Reserve Bank of St. Louis |
| DFII10 | Market Yield on U.S. Treasury Securities at 10-Year Constant Maturity, Inflation-Indexed, Percent, Daily, Not Seasonally Adjusted. | Federal Reserve Bank of St. Louis |
| CPIAUCSL | Consumer Price Index for All Urban Consumers: All Items in U.S. City Average, Index 1982-1984=100, Monthly, Seasonally Adjusted. Price index of a basket of goods and services paid by urban consumers. | Federal Reserve Bank of St. Louis |
| WILL5000PR | Wilshire 5000 Price Index, Index, Daily, Not Seasonally Adjusted. The Wilshire 5000 Total Market Index, or more simply the Wilshire 5000, is a market-capitalisation-weighted index of the market value of all American stocks actively traded in the United States. As of December 31, 2023, the index contained 3,403 components. | Federal Reserve Bank of St. Louis |
| Gold price | Gold price, USD per troy ounce, daily. | Bloomberg, Datastream, ICE Benchmark Administration, World Gold Council |
| *Dummy variables* | | |
| Dummy1 | Equals to 0 before 20.11. 2020 and 1 otherwise | Constructed by authors |
| Dummy2 | Equals to 0 before 8.11.2022 and 1 otherwise | Constructed by authors |
| Dummy3 | Equals to 1 between 20.11.2020 and 8.11.2022 and 0 otherwise | Constructed by authors |

All variables are treated as endogenous in the estimations, with the exception of *Total supply* (Table 2). The variable *Total supply* is considered to be exogenous because it is pre-determined by the Bitcoin algorithm, is publicly known and is expected to affect the dependent variable without being affected by it.[6]

To account for the structural change in the Bitcoin price development over time, we have included different dummy variables in the estimated models (Table 2). Dummy1 takes the value 0 before 20/11/2020 and 1 after this date and takes into account the implications of the Covid-19 pandemic related money and fiscal stimulus measure adopted by different countries. Dummy2 takes the value 0 before 8/11/2022 and 1 after this date, corresponding to the collapse of the FTX cryptocurrency exchange. First, we run estimations with these two dummy variables. For robustness, we rerun the models by considering Dummy3, which takes the value 1 between 20/11/2020 and 8/11/2022 and 0 outside this period.

**Table 3. Descriptive statistics**

| Variable | Obs | Mean | Std. Dev. | Min | Max |
| --- | --- | --- | --- | --- | --- |
| BTC price | 1514 | 28694.130 | 15231.340 | 5005.000 | 67547.000 |
| Bank netflow | 1514 | -0.002 | 1109.310 | -31329.600 | 3103.600 |
| Bank whale netflow | 1514 | -127.906 | 1129.314 | -32676.300 | 1973.000 |
| Bank reserve | 1514 | 2445.825 | 4932.195 | 0.000 | 41184.600 |
| Fund volume | 1514 | 1.77E+08 | 2.33E+08 | 1.48E+05 | 2.34E+09 |
| Exchange netflow | 1514 | -579.589 | 8200.169 | -69361.400 | 47550.600 |
| Exchange whale netflow | 1514 | -3901.359 | 7163.882 | -72858.800 | 39215.000 |
| Exchange reserve | 1514 | 2.60E+06 | 2.83E+05 | 2.06E+06 | 3.14E+06 |
| On chain BTC transactions | 1514 | 270196.600 | 103730.000 | -42081.600 | 711560.000 |
| Coin days destroyed | 1514 | 1.07E+07 | 1.23E+07 | 1.56E+06 | 1.99E+08 |
| Total supply | 1514 | 1.89E+07 | 4.11E+05 | 1.81E+07 | 1.96E+07 |
| DFF | 1514 | 1.845 | 2.125 | 0.040 | 5.330 |
| DFII10 | 1514 | 0.174 | 1.128 | -1.190 | 2.520 |
| CPIAUCSL | 1514 | 281.815 | 18.706 | 255.868 | 308.850 |
| WILL5000PR | 1514 | 40522.940 | 5529.191 | 22482.200 | 49252.260 |
| Gold price | 1514 | 1825.159 | 124.946 | 1459.650 | 2078.400 |

## 5 Results

### 5.1 Specification tests

Before estimating the ARDL model, it is essential to assess the stationarity of the series and determine their order of integration. Using the Augmented Dickey-Fuller (ADF) and Phillips-Perron (PP) tests, we reassured that none of the series is integrated of order I(2) or higher, as the ARDL methodology is not applicable in such cases.

After this step, we proceeded to test the existence of a long-term relationship between the time series. Since the null hypothesis of no long-term relationship was rejected in all specifications, we could estimate the error correction representation of the ARDL model[7]. Once the ARDL model is estimated, the short-term dynamics are captured by the lagged differences of the variables, while the long-term relationship is represented by the levels of the variables and the error correction term measures the speed at which the variables adjust towards equilibrium after a shock. As detailed in Table 5, the error correction terms as well as the long-term coefficients are nonzero, indicating the existence of a long-run relationship between variables. The long-term coefficients show the effect of a change in independent variables on the Bitcoin price in the long-run equilibrium.

---

[6] According to the Bitcoin algorithm, Bitcoins are created through a 'mining' process in which network participants use their computing power to verify and record payments on the blockchain. In return, they receive transaction fees and newly minted Bitcoins. After each block is created, a fixed number of Bitcoins are issued at a pre-determined and publicly known rate, increasing the total supply at a decreasing rate. This issuance rate is halved every four years and eventually converges to zero, capping the maximum supply at 21 million Bitcoins. Currently, miners receive 6.25 Bitcoins per block (approximately every 10 minutes).

[7] The results of unit root tests and ARDL bounds tests are available upon request from the authors.

In all model specifications, we employed robust standard errors to account for heteroscedasticity detected by the Breusch-Pagan test. Traditional standard errors may be biased in the presence of heteroscedasticity, potentially leading to incorrect inferences about the significance of coefficients (Pesaran et al. 2001). By using robust standard errors, we ensure that our hypotheses tests and confidence intervals are valid.

### 5.2 Set-up of Models M1 – M10

Table 4 summarises the estimated models. Models M1 to M7 differ by the alternative variables considered to capture off-chain and on-chain demand and supply drivers of Bitcoin transactions. This is done to account for possible correlations between different variables belonging to a given group of drivers, as well as for the fact that some variables from a given group may capture the same effect on the Bitcoin price. For example, Model M1 considers the variable *Bank netflow* for the off-chain demand side, *Exchange netflow* for the off-chain supply side, *On-chain BTC transactions* for the on-chain demand side and *Total supply* and *Coin days destroyed* for the on-chain supply side. Models M8 and M9 include only off-chain related variables, while model M10 includes only on-chain related variables. All estimated models include the same set of control variables related to the macro-financial environment, and two dummy variables related to the times series dimension (Dummy1 and Dummy2)

**Table 4. Specification of the estimated models**

| Variables | Hypothesis | M1 | M2 | M3 | M4 | M5 | M6 | M7 | M8 | M9 | M10 |
|---|---|---|---|---|---|---|---|---|---|---|---|
| *Off-chain demand* | | | | | | | | | | | |
| Bank netflow | H1 | x | | x | x | x | | | x | | |
| Bank whale netflow | H2 | | | | x | x | x | x | | | |
| Bank reserve | H1 | | x | | | | x | | | x | |
| Fund volume | H1 | | | x | | x | x | | x | x | |
| *Off-chain supply side* | | | | | | | | | | | |
| Exchange netflow | H1 | x | | x | x | x | | | x | | |
| Exchange whale netfl | H2 | | | | x | x | x | x | | | |
| Exchange reserve | H1 | | x | | | | x | | | x | |
| *On-chain demand* | | | | | | | | | | | |
| On-chain BTC transactions | H1 | x | x | x | x | x | x | | | | x |
| *On-chain supply side* | | | | | | | | | | | |
| Total supply | H1 | x | x | x | x | x | x | x | | | x |
| Coin days destroyed | H1 | x | x | x | x | x | x | x | | | x |
| *Macro-financial* | | | | | | | | | | | |
| DFF | | x | x | x | x | x | x | x | x | x | x |
| DFII10 | | x | x | x | x | x | x | x | x | x | x |
| CPIAUCSL | | x | x | x | x | x | x | x | x | x | x |
| WILL5000PR | | x | x | x | x | x | x | x | x | x | x |
| Gold price | | x | x | x | x | x | x | x | x | x | x |

The estimation results of the models are reported in Table 5 for long-run impacts and Table 6 for short-run impacts. The long-run coefficient estimates report whether the Bitcoin price is in a long-run equilibrium relationship with the considered covariates. The short-run impacts represent the immediate impact and short-run dynamics of the variables in the system, describing how the series react when the long-run equilibrium is disturbed. The estimation results of specifications with one dummy variable (Dummy3) present a robustness check are reported in Appendix Table 7 and Table 8.

**5.3 Hypothesis 1: Off-chain demand and supply drivers of Bitcoin price**

*Long-run impacts*

According to the results reported in Table 5, in all estimated models there are always several statistically significant off-chain and/or on-chain drivers exercising a long-run impact on the Bitcoin price. Regarding the off-chain-related variables, this is the case for *Bank netflow* and *Bank reserve*. These two variables are statistically significant in all models in which they were considered except for Model M5. The estimated coefficients associated with these variables have the expected sign: an increase in *Bank netflow* and *Bank reserve* exerts an upward pressure on the Bitcoin price. Variables *Bank whale netflow* and *Exchange whale netflow* are statistically significant in Model M6. The rest of off-chain-related variables are statistically not significant in the long-run relationship. Regarding the on-chain-related drivers, the variable *On chain BTC transactions* has a positive and statistically significant long-run impact on the Bitcoin price in all estimated models in which it was considered except in Models M5 and M6, which is in line with expectations. The variable *Coin days destroyed* is statistically not significant. The *Total supply* variable is considered exogenous and therefore has no long-run relationship with the Bitcoin price.

These results for long-run effects suggest that the demand-side drivers of Bitcoin transactions have a statistically significant and economically meaningful impact on the Bitcoin price. This is true for both off-chain and on-chain drivers. In contrast, off-chain and on-chain supply-side drivers are largely statistically not significant in most of the estimated models. Based on these results, we cannot reject hypothesis 1, which states that off-chain demand drivers have a significant impact on the Bitcoin price. Off-chain supply drivers are largely insignificant, implying that their impact on the Bitcoin price is negligible.

Our estimates also show that although off-chain transactions dominate the total Bitcoin activity, also on-chain transactions exert a statistically significant impact on the Bitcoin price. The level of activity on the blockchain reflects the network adoption of Bitcoin, representing its user base, trading activity, and the overall trust in the blockchain-based crypto economy. In other words, Bitcoin market fundamentals associated with the on-chain demand side are crucial drivers of its market valuation, as they encompass user engagement, transaction activity, and the confidence users have in the decentralised system. According to our estimates, they have a statistically significant impact on the Bitcoin price.

*Short-run impacts*

The short-run effects on the Bitcoin price are reported in Table 6. In contrast to the long-run relationship, both demand- and supply-side off-chain drivers affect the Bitcoin price in the short-run. All considered off-chain drivers ─ *Bank netflow*, *Bank whale netflow, Bank reserve*, *Fund volume*, *Exchange netflow*, *Exchange whale netflow* and *Exchange reserve* ─ are statistically significant in selected models. The demand-side off-chain drivers exert a more pronounced impact on the Bitcoin price than supply-side drivers. The demand-side off-chain drivers are statistically significant in all estimated models in the short-run, while the off-chain supply drivers are statistically significant in most Models (M2 to M6, M8 and M9 in Table 6). Further, our estimates show that the off-chain demand drivers have statistically significant short-run effects up to the third lag, while the off-chain supply drivers have statistically significant short-run effects only up to the first lag. This suggests that while the demand drivers have an impact over a longer period of time, the impact of off-chain supply-side drivers diminishes faster.

Regarding the on-chain-related variables, only the demand-side driver is statistically significant – similar to long-run impacts. Second, on-chain demand-side drivers are statistically significant in fewer Models (M2 to M6 in Table 6) compared to the long-run estimates. Further, in the short-run, only the contemporaneous coefficients are statistically significant, indicating that the immediate effects of on-chain demand are more pronounced than those in subsequent periods (days).

Overall, based on the estimated short-run effects, we cannot reject Hypothesis 1. Both off-chain demand- and supply-side drivers significantly affect the Bitcoin price in the short-run. In contrast, on-chain drivers exert a pronounced impact, suggesting that off-chain factors dominate the relationship to the Bitcoin price in the short-run. Our estimates also confirm structural differences between on- and off-chain drivers of the Bitcoin price. The difference between short- and long-run results can be explained by several factors. As off-chain transactions mainly take place on CEXs, they are mainly associated with speculative investments aimed at generating profits from short-term price movements, rather than directly participating in the economic activity (market fundamentals), such as the purchase of goods and services (Ciaian et al. 2016; Kukacka and Kristoufek 2023). This is confirmed by estimates in Table 5 and Table 6, where the short-term speculative effects of off-chain transactions dominate the long-term off-chain effects on the Bitcoin price. In contrast, on-chain Bitcoin trading (e.g. on DEXs) is significantly less common. Therefore, on-chain Bitcoin transactions are more likely to be driven by market fundamentals in the long-run, and thus impact the Bitcoin price over longer time period.

Our estimates indeed suggest that on-chain drivers affect Bitcoin price through different channels compared to off-chain drivers. These findings are consistent with the literature that emphasises the dual nature of the Bitcoin price dynamics, where both investor speculative behaviour as well as market fundamentals drive the Bitcoin price (Yae and Tian 2024). In contrast, our findings contradict studies that primarily emphasise the role of speculative investments in driving the Bitcoin price (Kukacka and Kristoufek 2023). Our paper contributes to this literature by separately identifying short-term and long-term Bitcoin price drivers of on- and off-chain transactions: short-term Bitcoin price movements are mostly driven by speculative factors, while longer-term Bitcoin price movements are largely influenced by market fundamentals.

### 5.4 Hypothesis 2: "Whales" drivers of the Bitcoin price

*Long-run impacts*

We investigate if and how cryptocurrency whales influence market trends with substantial transactions. According to the results reported in Table 5, the considered off-chain whale demand and supply drivers of Bitcoin transactions (*Bank whale netflow* and *Exchange whale netflow*) are statistically significant in Models M6 and M7. As expected, *Bank whale netflow* variable has a positive impact on the Bitcoin price, while *Exchange whale netflow* has a negative impact on the Bitcoin price. These results weakly support Hypothesis, 2 suggesting that the trading patterns of whales may have a limited impact on the Bitcoin price. However, a further research using more nuanced data on whale transactions is required to provide a definite answer regarding crypto-whale trading – Bitcoin price relationship. For example, extending the whale cohort to top 1,000 holders or top 10,000 holders and reassessing Hypothesis 2 offers a promising avenue for future research.

*Short-run impacts*

In contrast to the long-run effects, the short-run effects of whales' trading are more pronounced on the Bitcoin price (Table 6). In all estimated models one or more whale related variables are statistically significant. Specifically, while the included off-chain whale demand driver of Bitcoin transactions is statistically significant in Models M6 and M7 (*Bank whale netflow*) – as in the case of long-run estimates – the off-chain whale supply driver (*Exchange whale netflow*) is statistically significant in more estimated models in the short-run (Models M4, M5 and M7) than in the long-run (Model M6). Based on these short-run estimates we cannot reject Hypothesis 2, postulating that whale trading patterns affect the price of Bitcoin in the short-run.

Overall, the estimation results for Hypothesis 2 suggest that the whale trading can influence crypto-market supply and demand and cause Bitcoin price movements, and/or that their trading can strategically exploit market inefficiencies in the short term. However, the whale trading behaviour is found to be less significant for the long-term price movements of Bitcoin. These findings support the well-established evidence in the financial literature that large traders – whales – can exploit market inefficiencies and cause short-term price volatility (Gabaix et al. 2006; Choi and Chhabria 2012; Merkley et al. 2024). This result is also consistent with the evidence arguing that large Bitcoin trades are driving significant price movements. Despite ongoing the ongoing speculations in mass media, this question is little explored in the empirical Bitcoin literature statistically, and offers a promising avenue for future research.

**5.5 Macro-financial drivers of the Bitcoin price**

*Long-run impacts*

Our estimation results suggest a somewhat weaker dependence of the Bitcoin price on macro-financial developments in the long-run. Most of the macro-financial variables considered are statistically not significant in most of the estimated models (Table 5). The exceptions are variables *CPIAUCSL* (Consumer Price Index) in Models M2 to M6, *DFF* (Federal Funds Effective Rate) in Model M6 and *Gold price* in Model M9. The remaining macro-financial variables do not exercise a statistically significant impact on the Bitcoin price in the long-run.

Among macro-financial variables, the Bitcoin price seems most affected by the inflationary pressures and monetary policy. In line with Cong et al. (2024), the *DFF* variable is negatively related to the Bitcoin price in the long-run, suggesting that the Bitcoin price decreases when monetary policy is contractionary and increases when it is expansionary. Contrary to expectations, *CPIAUCSL* has a negative long-run impact on the price of Bitcoin in Models M2 to M6, suggesting that Bitcoin's role as a store of value is rather limited. In contrast, it is likely that Bitcoin is perceived as an investment asset, so that a contractionary monetary policy implemented by National Banks in times of high inflation may reduce liquidity in markets and cause asset prices, including Bitcoin, to fall (Sören 2023; Apergis 2024). This is further supported by the negative relationship estimated between the *Gold price* and the Bitcoin price. This estimation result for *Gold price* suggests that Bitcoin may not be in a competitive relationship with gold as an alternative investment or as a store of value during the four-year period considered (2020-2024). The previous evidence in the literature is mixed (Apergis 2024; Ozer et al. 2024).

*Short-run impacts*

Compared to long-run estimates, macro-financial variables tend to have a greater impact on the Bitcoin price in the short-run. Although, most macro-financial variables are not statistically significant in most estimated models in the short-run, few variables are statistically significant in all models. The macro-financial variables with the highest significance level are *WILL5000PR* (American stocks actively) in all estimated models and *CPIAUCSL* (Consumer Price Index) in Models M2 to M6. The stock market performance (*WILL5000PR*) affects the

Bitcoin price both immediately (contemporaneously) and in the short term (first and second lags). This suggests that the impact of the stock market performance on the Bitcoin price is temporarily dissipating, before the market returns to its long-run equilibrium. Further, *CPIAUCSL* is also found to have a contemporaneous impact on the Bitcoin price though its role in the short-run appears to be similar as in the long-run. This is evidenced by its significance in the same models and at a lower level of significance in both short- and long-run. The rest of macro-financial variables are not statistically significant in the short-run.

Overall, whereas macro-financial variables have a limited impact on the Bitcoin price in the long-run, their influence seems to be more pronounced in the short-run. While the inflation variable (*CPIAUCSL*) has an impact on the Bitcoin price both in the short- and long-run, the stock market performance (*WILL5000PR*) strongly influences the Bitcoin price in the short term. This reinforces the view that Bitcoin operates as a speculative asset rather than a safe haven during market volatility periods, responding to both macroeconomic indicators and short-term financial dynamics. Our findings are consistent with the literature estimating a differentiated temporal impact of macro-financial drivers on the Bitcoin price and highlight the Bitcoin's sensitivity to immediate market sentiment, inflation expectations and investor behaviour (Ciaian 2016; Cong et al. 2024).

## 5.6 Robustness analyses

The robustness estimates that consider one dummy variable in Table 7 and Table 8 largely confirm the results presented in the previous section in Table 5 and Table 6. Similar to the main results, the long-run robustness estimates are partially in line with Hypothesis 1: off-chain demand drivers exert a significant impact on the Bitcoin price, whereas off-chain supply drivers remain largely insignificant. Furthermore, the robustness results are consistent with Hypothesis 1 regarding the short-run effects, showing that both off-chain demand and supply drivers significantly affect the Bitcoin price in the short-run.

Contrary to the main results, the robustness estimates suggest a dampened impact of on-chain demand drivers on the Bitcoin price in the long-run, while indicating a more pronounced effect in the short-run. These results suggest that on-chain factors have some impact on the Bitcoin price in the long-run, while also exert a significant impact on the dynamics of the Bitcoin price in the short-run. Overall, the robustness estimates are consistent with the main results for Hypothesis 2 in terms of both short- and long-run effects. Specifically, they partially confirm Hypothesis 2 in the long-run and are fully in-line with it in the short-run. Whale trading patterns have a weak effect on the Bitcoin price in the long-run, but a significant effect in the short-run.

**Table 5. Estimation results: long-run impacts (models with two dummies)**

|                          | M1         | M2         | M3         | M4         | M5         | M6          | M7         | M8         | M9         | M10        |
|--------------------------|------------|------------|------------|------------|------------|-------------|------------|------------|------------|------------|
| Bank netflow             | 0.112 *    |            | 0.148 ***  | 0.352      | 0.459      |             |            | 0.108 *    |            |            |
| Bank reserve             |            | 0.024 **   |            |            |            | 0.030 ***   |            |            | 0.028 ***  |            |
| Bank whale netflow       |            |            |            | -0.201     | -0.305     | 0.089 *     | 0.096 *    |            |            |            |
| Fund volume              |            |            | 0.000      |            | 0.000      | 0.000       |            | 0.000      | 0.000      |            |
| Exchange netflow         | 0.003      |            | 0.003      | 0.005      | 0.006      |             |            | 0.002      |            |            |
| Exchange reserve         |            | 0.000      |            |            |            | 0.000       |            |            | 0.000      |            |
| Exchange whale netflow   |            |            |            | -0.001     | -0.002     | -0.018 **   | 0.003      |            |            |            |
| On chain BTC transactions| 0.001 *    | 0.001 *    | 0.001 *    | 0.001 *    | 0.001      | 0.001       |            |            |            | 0.001 ***  |
| Coin days destroyed      | 0.000      | 0.000      | 0.000      | 0.000      | 0.000      | 0.000       | 0.000      | 0.000      |            | 0.000      |
| DFF                      | -53.855    | -111.971   | -55.003    | -79.740    | -76.018    | -124.139 *  | 22.570     | 9.269      | -38.785    | -86.920    |
| DFII10                   | -127.657   | -164.505   | -115.543   | -138.047   | -167.477   | -191.132    | -108.405   | -136.415   | -212.709   | -124.943   |
| CPIAUCSL                 | -43.470    | -53.606 *  | -53.220 *  | -58.007 *  | -58.610 *  | -58.914 *   | -21.613    | 2.332      | 9.332      | -44.530    |
| WILL5000PR               | -0.005     | 0.001      | -0.008     | -0.003     | -0.007     | -0.006      | 0.005      | 0.003      | 0.008      | -0.003     |
| Gold price               | -0.600     | -0.707     | -0.543     | -0.595     | -0.619     | -0.779      | -0.316     | -0.365     | -0.957 **  | -0.679     |
| Error correction term    |            |            |            |            |            |             |            |            |            |            |
| BTC price (-1)           | -0.014 **  | -0.016 *** | -0.013 **  | -0.016 *** | -0.014 **  | -0.014 **   | -0.012 **  | -0.011 **  | -0.011 **  | -0.017 *** |

**Table 6. Estimation results: short-run impacts (models with two dummies)**

| | M1 | M2 | M3 | M4 | M5 | M6 | M7 | M8 | M9 | M10 |
|---|---|---|---|---|---|---|---|---|---|---|
| Δ BTC price (-1) | 0.155 *** | 0.153 *** | 0.160 *** | 0.157 *** | 0.148 *** | 0.137 *** | 0.165 *** | 0.164 *** | 0.157 *** | 0.165 *** |
| Δ Bank netflow | -0.119 ** | | -0.154 *** | -0.606 * | -0.602 * | | | -0.115 ** | | |
| Δ Bank netflow (-1) | -0.030 | | -0.066 ** | -0.068 * | -0.076 ** | | | -0.027 | | |
| Δ Bank netflow (-2) | 0.082 *** | | 0.044 *** | 0.051 *** | 0.042 *** | | | 0.083 *** | | |
| Δ Bank netflow (-3) | 0.031 | | | | | | | 0.036 | | |
| Δ Bank reserve | | -0.019 | | | | | | | -0.021 | |
| Δ Bank reserve (-1) | | 0.077 *** | | | | | | | 0.075 *** | |
| Δ Bank reserve (-2) | | 0.098 *** | | | | | | | 0.094 *** | |
| Δ Bank reserve (-3) | | -0.062 *** | | | | | | | -0.059 *** | |
| Δ Bank whale netflow | | | | 0.451 | 0.442 | -0.116 *** | -0.109 ** | | | |
| Δ Bank whale netflow (-1) | | | | | | -0.047 | -0.023 | | | |
| Δ Bank whale netflow (-2) | | | | | | 0.053 *** | 0.093 *** | | | |
| Δ Bank whale netflow (-3) | | | | | | | 0.038 | | | |
| Δ Fund volume | | | 0.000 | | 0.000 | 0.000 | | 0.000 | 0.000 | |
| Δ Fund volume (-1) | | | | | | | | 0.000 | 0.000 | |
| Δ Fund volume (-2) | | | | | | | | 0.000 | 0.000 | |
| Δ Fund volume (-3) | | | | | | | | 0.000 | 0.000 * | |
| Δ Exchange netflow | 0.005 | | 0.005 * | 0.014 ** | 0.015 ** | | | 0.006 * | | |
| Δ Exchange netflow (-1) | | | | 0.009 | 0.009 | | | | | |
| Δ Exchange reserve | | 0.007 * | | | | 0.021 *** | | | 0.008 ** | |
| Δ Exchange reserve (-1) | | -0.005 | | | | -0.005 | | | -0.006 * | |
| Δ Exchange whale netflow | | | | -0.015 * | -0.016 ** | | 0.000 | | | |
| Δ Exchange whale netflow (-1) | | | | -0.015 ** | -0.015 ** | | -0.006 ** | | | |
| Δ On chain BTC transactions | 0.001 | 0.001 * | 0.001 * | 0.001 * | 0.001 * | 0.001 ** | | | | |
| Δ On chain BTC transactions (-1) | 0.001 | 0.001 | 0.001 | 0.001 | 0.001 | 0.001 | | | | |
| Δ CPIAUCSL | 241.999 | 256.053 * | 248.526 | 260.936 * | 262.495 * | 265.511 * | 233.337 | 211.881 | 213.617 | 246.179 |
| Δ CPIAUCSL (-1) | 40.051 | 50.772 | | | | | 17.022 | 13.547 | 14.006 | 27.703 |
| Δ CPIAUCSL (-2) | -3.837 | 8.474 | | | | | -4.229 | -12.210 | -12.588 | -0.379 |
| Δ CPIAUCSL (-3) | -221.316 | -209.856 | | | | | -232.355 | -254.602 | -253.482 | -235.642 |
| Δ WILL500PR | 0.390 *** | 0.382 *** | 0.377 *** | 0.376 *** | 0.372 *** | 0.379 *** | 0.389 *** | 0.393 *** | 0.386 *** | 0.385 *** |
| Δ WILL500PR (-1) | 0.353 *** | 0.356 *** | 0.354 *** | 0.359 *** | 0.373 *** | 0.374 *** | 0.362 *** | 0.366 *** | 0.371 *** | 0.352 *** |
| Δ WILL500PR (-2) | -0.085 * | -0.085 * | -0.096 * | -0.094 * | | | -0.080 | -0.083 * | -0.081 * | -0.083 * |
| Total supply | -0.005 * | -0.006 ** | -0.005 ** | -0.006 ** | -0.007 ** | -0.007 ** | 0.000 | | | -0.005 ** |
| dummy1 | 6033.423 | -31854.990 | 2272.856 | -13517.150 | -11578.600 | -28406.550 | -9403.161 | 12044.270 | 7721.753 | 5327.651 |
| dummy2 | 3464.038 | 18935.180 | 7046.416 | 16743.350 | 10875.930 | 15096.870 | -16701.15 | -36602.710 *** | -39669.350 *** | 7389.219 |
| timedummy1 | -0.260 | 1.435 | -0.091 | 0.614 | 0.527 | 1.281 | 0.443 | -0.520 | -0.337 0.543 | -0.225 |
| timedummy2 | -0.149 | -0.823 | -0.306 | -0.727 | -0.472 | -0.654 | 0.726 | 1.597 *** | 1.731 *** | -0.318 |
| Trend | 7.009 ** | 7.326 * | 7.948 *** | 8.341 *** | 9.015 *** | 9.040 ** | | | | 7.320 ** |
| Const | -54786.400 *** | -41939.440 | -60441.380 *** | -58710.210 *** | -63095.020 *** | -53227.820 | | | | -56297.360 *** |

# 6 Conclusions

This paper investigates the impact of different types of Bitcoin transactions (on-chain versus off-chain) on the Bitcoin price over the short and long term. After examining Bitcoin transaction patterns, we develop two hypotheses regarding the influence of on-chain and off-chain demand and supply factors on the Bitcoin price. Hypothesis 1 posits that the price of Bitcoin is predominantly influenced by off-chain demand and supply drivers. Hypothesis 2 states that trading behaviours of large Bitcoin traders ("whales") have a significant impact on Bitcoin price. As usual, we include controls for macro-financial developments. To investigate these two questions, we apply time-series analytical mechanisms to daily data from 2019 to 2024.

Our findings partially confirm Hypothesis 1 in the long-run suggesting that off-chain demand drivers have a significant impact on the Bitcoin price, whereas off-chain supply drivers do not. In the short-run, both off-chain demand and supply drivers significantly affect the Bitcoin price, thus fully in line with Hypothesis 1. These results provide support to the evidence that speculative drivers dominate the Bitcoin price formation, as off-chain transactions take place mainly on CEXs and are mainly associated with speculative investor trading aimed at generating profits from short-term price movements, rather than directly supporting economic activity (market fundamentals) such as the purchase of goods and services. Regarding transactions on the blockchain, on-chain demand drivers tend to affect the Bitcoin price in the long-run, while both on-chain demand and supply drivers affect the Bitcoin price in the short-run. These findings point to the dual nature of the Bitcoin price dynamics, where market fundamentals affect Bitcoin prices in addition to the investor speculative behaviour stated by Hypothesis 1.

Our results suggest that whale trading patterns have a weak impact on the Bitcoin price in the long-run, but a more pronounced impact in the short-run. This confirms the differential impact of Hypothesis 2 across time dimensions, which is consistent with the pattern observed in Hypothesis 1. These results imply that a whale trading can influence the supply and demand for Bitcoin and cause price fluctuations, especially in the shorter time perspective. Moreover, their strategic trading behaviour appears to be adept at exploiting market inefficiencies during high-volatility periods.

In terms of macro-financial factors, we find that Bitcoin's price is primarily affected by inflationary pressures and monetary policy in the long-run. Surprisingly, inflation has a negative impact in the long-run, suggesting that Bitcoin plays a limited role as a hedge during inflation. In the short term, macro-financial variables are found to have greater impact on the Bitcoin price, driven by inflationary pressures and especially stock market developments. This reinforces the view that Bitcoin acts as a speculative asset rather than a safe harbour during periods of high market volatility.

Our estimates confirm that on-chain drivers affect Bitcoin price through different channels compared to off-chain drivers. Our paper contributes to this literature by separately identifying short-term and long-term Bitcoin price drivers of on-and off-chain transactions: short-term Bitcoin price movements are mostly driven by speculative factors, while longer-term Bitcoin price movements are largely influenced by market fundamentals.

As discussions surrounding the crypto-asset distribution continue both among policy makers and investors, many stakeholders are looking to the future of cryptocurrencies and the evolving landscape of crypto-assets. Efforts to address this concentration issue might include implementing policies that encourage broader participation, promoting financial literacy, and creating mechanisms that incentivise long-term holding over short-term speculations.

# Appendix: Additional tables

## Table 7. Estimation results: long-run impacts (models with one dummy)

| | M1 | M2 | M3 | M4 | M5 | M6 | M7 | M8 | M9 | M10 |
|---|---|---|---|---|---|---|---|---|---|---|
| Bank netflow | 0.108 * | | 0.146 *** | 0.340 | 0.449 | | | 0.113 ** | | |
| Bank reserve | | 0.023 ** | | | | 0.029 *** | | | 0.026 ** | |
| Bank whale netflow | | | | -0.190 | -0.295 | 0.093 ** | 0.100 * | | | |
| Fund volume | | | 0.000 | | 0.000 | 0.000 | | 0.000 | 0.000 | |
| Exchange netflow | 0.003 | | 0.003 | 0.005 | 0.006 | | | 0.003 | | |
| Exchange reserve | | 0.000 | | | | 0.000 | | | 0.000 * | |
| Exchange whale netflow | | | | -0.001 | -0.002 | -0.018 ** | 0.003 | | | |
| On chain BTC transactions | 0.001 *** | 0.001 | 0.001 | 0.001 | 0.001 | 0.001 | | | | 0.001 ** |
| Coin days destroyed | 0.000 | 0.000 | 0.000 | 0.000 | 0.000 | 0.000 | 0.000 | | | 0.000 |
| DFF | -37.485 | -95.512 | -37.635 | -57.673 | -60.497 | -109.109 * | -1.506 | 16.903 | -32.944 | -59.286 |
| DFII10 | -100.325 | -198.029 | -160.766 | -192.112 | -203.931 | -220.256 * | -134.869 | 7.648 | -11.122 | -188.540 |
| CPIAUCSL | -5.392 | -31.017 * | -38.679 ** | -36.103 ** | -42.447 ** | -39.276 ** | -40.573 *** | -1.417 | 2.524 | -33.766 ** |
| WILL5000PR | 0.010 | -0.001 | -0.009 | -0.004 | -0.009 | -0.009 | -0.008 | 0.014 | 0.022 | -0.006 |
| Gold price | -0.853 ** | -0.958 ** | -0.690 | -0.828 * | -0.795 * | -1.008 ** | -0.389 | 0.011 | -0.374 | -0.833 * |
| Error correction term | | | | | | | | | | |
| BTC price (-1) | -0.014 *** | -0.015 *** | -0.013 ** | -0.015 *** | -0.014 *** | -0.013 ** | -0.012 ** | -0.009 * | -0.009 * | -0.017 *** |

**Table 8. Estimation results: short-run impacts (models with one dummy)**

| | M1 | | M2 | | M3 | | M4 | | M5 | | M6 | | M7 | | M8 | | M9 | | M10 | |
|---|---|---|---|---|---|---|---|---|---|---|---|---|---|---|---|---|---|---|---|---|
| Δ BTC price (-1) | 0.157 | *** | 0.153 | *** | 0.160 | *** | 0.157 | *** | 0.148 | *** | 0.137 | *** | 0.163 | *** | 0.165 | *** | 0.159 | *** | 0.167 | *** |
| Δ Bank netflow | -0.114 | ** | | | -0.153 | *** | -0.600 | * | -0.596 | * | | | | | -0.120 | ** | | | | |
| Δ Bank netflow (-1) | -0.025 | | | | -0.065 | ** | -0.067 | * | -0.076 | ** | | | | | -0.030 | | | | | |
| Δ Bank netflow (-2) | 0.085 | *** | | | 0.045 | *** | 0.051 | *** | 0.042 | *** | | | | | 0.081 | *** | | | | |
| Δ Bank netflow (-3) | 0.032 | | | | | | | | | | | | | | 0.035 | | | | | |
| Δ Bank reserve | | | -0.018 | | | | | | | | | | | | | | -0.018 | | | |
| Δ Bank reserve (-1) | | | 0.078 | *** | | | | | | | | | | | | | 0.078 | *** | | |
| Δ Bank reserve (-2) | | | 0.099 | *** | | | | | | | | | | | | | 0.097 | *** | | |
| Δ Bank reserve (-3) | | | -0.061 | *** | | | | | | | | | | | | | -0.057 | *** | | |
| Δ Bank whale netflow | | | | | | | 0.445 | | 0.436 | | -0.118 | *** | -0.113 | ** | | | | | | |
| Δ Bank whale netflow (-1) | | | | | | | | | | | -0.048 | | -0.026 | | | | | | | |
| Δ Bank whale netflow (-2) | | | | | | | | | | | 0.052 | *** | 0.091 | *** | | | | | | |
| Δ Bank whale netflow (-3) | | | | | | | | | | | | | 0.036 | | | | | | | |
| Δ Fund volume | | | | | 0.000 | | | | 0.000 | | 0.000 | | | | 0.000 | | 0.000 | | | |
| Δ Fund volume (-1) | | | | | | | | | | | | | | | 0.000 | | 0.000 | | | |
| Δ Fund volume (-2) | | | | | | | | | | | | | | | 0.000 | | 0.000 | | | |
| Δ Fund volume (-3) | | | | | | | | | | | | | | | 0.000 | | 0.000 | * | | |
| Δ Exchange netflow | 0.005 | | | | 0.005 | * | 0.014 | ** | 0.015 | ** | | | | | 0.006 | * | | | | |
| Δ Exchange netflow (-1) | | | | | | | 0.009 | | 0.009 | | | | | | | | | | | |
| Δ Exchange reserve | | | 0.007 | * | | | | | | | 0.021 | *** | | | | | 0.009 | ** | | |
| Δ Exchange reserve (-1) | | | -0.005 | | | | | | | | -0.005 | | | | | | -0.006 | * | | |
| Δ Exchange whale netflow | | | | | | | -0.015 | * | -0.016 | ** | | | 0.000 | | | | | | | |
| Δ Exchange whale netflow (-1) | | | | | | | -0.016 | ** | -0.015 | ** | | | | | -0.006 | * | | | | |
| Δ On chain BTC transactions | | | 0.001 | * | 0.001 | ** | 0.001 | * | 0.001 | ** | 0.001 | ** | | | | | | | 0.001 | |
| Δ On chain BTC transactions (-1) | | | 0.001 | * | 0.001 | | 0.001 | * | 0.001 | * | 0.001 | * | | | | | | | | |
| Δ DFF | | | | | | | | | | | | | | | | | | | 310.282 | |
| Δ CPIAUCSL | 218.307 | | 245.796 | | 242.628 | | 252.131 | * | 255.642 | * | 257.029 | * | 246.045 | | 209.395 | | 210.606 | | 249.117 | |
| Δ CPIAUCSL (-1) | 17.692 | | 40.085 | | | | | | | | | | 28.827 | | 10.269 | | 10.020 | | 30.044 | |
| Δ CPIAUCSL (-2) | -21.075 | | -2.431 | | | | | | | | | | 6.428 | | -14.011 | | -14.160 | | -0.920 | |
| Δ CPIAUCSL (-3) | -245.583 | | -219.705 | | | | | | | | | | -222.074 | | -257.168 | | -256.263 | | | |
| Δ WILL500PR | 0.389 | *** | 0.384 | *** | 0.378 | *** | 0.377 | *** | 0.374 | *** | 0.381 | *** | 0.393 | *** | 0.389 | *** | 0.382 | *** | 0.382 | *** |
| Δ WILL500PR (-1) | 0.356 | *** | 0.358 | *** | 0.355 | *** | 0.360 | *** | 0.373 | *** | 0.376 | *** | 0.366 | *** | 0.364 | *** | 0.368 | *** | 0.351 | *** |
| Δ WILL500PR (-2) | -0.086 | * | -0.082 | * | -0.095 | ** | -0.092 | * | | | -0.077 | | -0.087 | * | -0.086 | * | -0.084 | * | | |
| Total supply | 0.001 | ** | -0.005 | ** | -0.004 | * | -0.005 | * | -0.006 | ** | -0.007 | *** | -0.004 | * | | | -0.004 | * | | |
| dummy3 | 13647.610 | ** | 3482.865 | | 9471.468 | | 6231.539 | | 5906.028 | | 5117.300 | | 9219.266 | | 14790.760 | ** | 8968.752 | | 12240.960 | ** |
| Timedummy3 | -0.599 | ** | -0.156 | | -0.413 | | -0.274 | | -0.259 | | -0.229 | | -0.401 | | -0.648 | ** | -0.397 | | -0.536 | ** |
| Trend | | | 7.328 | ** | 6.337 | ** | 6.885 | ** | 8.244 | ** | 9.309 | *** | 6.103 | * | | | | | 6.318 | ** |
| Const | -11619.32 | *** | -54930.09 | ** | -51470.28 | ** | -54714.330 | *** | -63121.570 | *** | -67544.69 | *** | -47983.48 | ** | | | | | -51576.37 | ** |